\documentclass[12pt]{article}
\begin{document}

\title{Dark Energy in an Axion Model with Explicit Z(N) Symmetry
Breaking}

\author{Pankaj Jain\\
Physics Department, IIT, Kanpur - 208016, India\\
}
\maketitle

\abstract{\small We point out that a well known axion model with 
an explicit Z(N) symmetry breaking term predicts both dark energy
and cold dark matter. We estimate the parameters of this model which
fit the observed densities of the dark components of the universe. We 
find that the parameters do not conflict with any observations.}

\bigskip

In this paper we point out that a well known axion action, originally
proposed by Sikivie \cite{sikivie82}, may account for both the dark energy
and dark matter present in the universe. Axion is a pseudo Goldstone 
boson due to the anomalous nature of the Peccei-Quinn symmetry 
$U_{\rm PQ}(1)$ \cite{PQ,weinberg,Dine}. 
QCD instanton effects generate a mass for the axion.
As pointed out in Ref. \cite{sikivie82} a discrete Z(N) subgroup of  
$U_{\rm PQ}(1)$ remains unbroken. 
Hence the effective potential experienced by the axion has N degenerate
vacua at $\theta = Na/f_{\rm PQ} = 2n\pi$, where $n=0, 1, ...(N-1)$, 
$a$ is the axion field
and $f_{\rm PQ}$ the scale of Peccei-Quinn symmetry breaking. 
In the early universe the scalar field can take any value. As the universe
cools the axion must settle into one of the minima of its potential.
Since there exist several minima for $N>1$, we expect the formation
of domain walls separating regions where axion takes different values.
These domain walls can considerably modify the cosmological evolution
of the universe and hence may conflict with observations. 

In the model proposed by Sikivie, there also exists an explicit 
$U_{\rm PQ}(1)$  symmetry breaking term. This term was originally introduced
so that the domain walls may decay at some early time 
and hence do not dominate the present energy density of  
the universe. The parameter range over which such a solution can work is
considerably restricted, although not entirely ruled out \cite{chang}. 
Here we avoid this problem by assuming that 
the universe inflated before or during PQ symmetry breaking. 
In this case the axion field is uniform over the observable
universe \cite{KolbTurner}. 
We may speculate that the spontaneous PQ symmetry breaking
may itself be responsible for inflation in the early universe. 

The explicit PQ symmetry breaking term can give rise to vacuum energy
at the current time if the
value of the axion field in the observable universe does not correspond to 
its true vacuum. In this case the universe is trapped in a false vacuum
and can provide a reasonable solution to the dark energy problem provided the 
lifetime of the false vacuum is larger than the age of the universe. 
In this paper we show that this is indeed true.

The effective lagrangian for the axion may be taken to be of form
\cite{sikivie82,chang}
\begin{equation}
{\cal L} = {1\over 2} \partial_\mu a\partial^\mu a + {m_a^2 f_{\rm PQ}^2\over
N^2} \left[\cos(Na/f_{\rm PQ}) -1\right]
\end{equation}
where $m_a$ is the mass of the axion.
The symmetry breaking term takes the form
\begin{equation}
\delta{\cal L} = \xi(\Phi e^{-i\delta} + h.c.) - {\cal E}_{\rm vac} 
\end{equation}
where $\Phi = f_{\rm PQ} e^{ia/f_{\rm PQ}}$ and $\xi$, 
$\delta$ are parameters. 
Here we have included a constant term  $ {\cal E}_{\rm vac}$
in order that the minimum value of
the potential is zero. This amounts to the assumption that the lowest energy 
state, or the true vacuum, has exactly zero energy. 
The symmetry breaking term can also be written as,
\begin{equation}
\delta{\cal L} = 2\xi f_{\rm PQ} \cos\left({a\over f_{\rm PQ}}
-\delta\right) - {\cal E}_{\rm vac} 
\end{equation}
The false vacuum gives rise to dark energy density of order
\begin{equation}
\delta {\cal E} \approx \xi f_{\rm PQ} 
\end{equation}
We set this equal to the observed dark energy density $(0.003\ {\rm eV})^4$ 
to get an estimate
for $\xi$,
\begin{equation}
\xi \approx (0.003\ {\rm eV})^4/f_{\rm PQ} = 8.1\times 10^{-27}\ {\rm eV}^3\  
\left({10^7\ {\rm GeV}
\over f_{\rm PQ}}\right)\ .
\end{equation}

This symmetry breaking term also produces an
effective value for the QCD $\theta$ parameter, $\theta_{\rm QCD}\approx \xi/ 
m_a^2 f_{\rm PQ}$. The axion mass can be expressed as, see for example
\cite{KolbTurner},
\begin{equation}
m_a = 0.62\ {\rm eV}\ \left({10^7\ {\rm GeV}\over f_{\rm PQ}/N} \right)
\end{equation}
We can, therefore, express $\theta_{\rm QCD}$ as,
\begin{equation}
\theta_{\rm QCD} \approx {\delta {\cal E}\over m_a^2 f_{\rm PQ}^2}
\approx {2.1\times 10^{-42}\over N^2}  \ ,
\end{equation}
which is much smaller compared to the current experimental limits. 

We next estimate the lifetime of such a false vacuum. 
For this we estimate the probability of bubble nucleation which is given
by \cite{coleman,KolbTurner}
\begin{equation}
\Gamma = A e^{-S_E}\ .
\end{equation}
Here $A$ can be estimated by dimensional analysis and 
$S_E$ is the Euclidean action. The factor $S_E$ is given by
\begin{equation}
S_E \approx {27\pi^2 \left[\int_0^{a_0} da\sqrt{2V_0(a)}\right]^4
\over 2(\Delta V)^3}\ 
\end{equation}
where $V_0$ is the effective potential ignoring the symmetry breaking
term and $\Delta V$ is the difference in the energy densities between
the false and the true vacua. We have taken these to be two adjacent 
minima such that the axion field at the false and true vacua is equal to 
$\phi_0$ and 0 respectively. Using the results above, we estimate this
factor to be
\begin{equation}
S_E \approx {4\times 10^{162} \over N^4} \left({f_{\rm PQ}\over 10^7\
{\rm GeV}}\right)^4\ .
\end{equation}
It is, therefore, clear that for any acceptable value of $f_{\rm PQ}$,
$\exp(-S_E)$ is an extraordinarily small number. Hence the probability
of transition from the false vacuum is extremely small and the false vacuum
will survive much longer than the age of the universe. 

Finally we point that this model 
also provides a candidate for the 
cold dark matter in the universe in the form of oscillations
of the axion condensate about its local minimum \cite{KolbTurner}. 
For this to provide an
appreciable component of the cold dark matter we require 
$m_a$ of order $10^{-5}$ eV, which implies $f_{\rm PQ}$ of order 
$10^{12}$ GeV.

Hence we find that this well known axion model can account for both
the cold dark matter as well as the dark energy present in the universe. 
The model is interesting since it does not introduce any new particles,
besides the axion.
The model requires a symmetry breaking term with a mass scale of order 
$0.003$ eV, as required by cosmological observations. We do not provide
any justification for introducing such a low mass scale except to state
that it is comparable to the scale of neutrino masses \cite{neutrino}.
Furthermore so far we do not
have any observable consequence of this proposal. The QCD $\theta$ parameter
turns out to be extremely small. Fortunately the lifetime of the
false vacuum also turns out to be very large. The proposal may be testable 
by fixing the axion parameters through its interaction with
other fields such as the photon \cite{Jain04}. For this we need to study 
their propagation in regions of strong magnetic fields such as in the pulsar 
magnetosphere. The mixing of axions with photons is negligible for galactic or 
intergalactic propagation if we assume the $PQ$ symmetry breaking scale of
order $10^{12}$ GeV. Axions and other light
pseudoscalars might also provide the cosmological seed magnetic fields due to 
their coupling to electromagnetic fields \cite{Garretson,Ng}. 
It is also possible that 
the symmetry breaking term discussed in this paper, or its generalization,
might have other observable predictions. The possibility that dark energy
may be explained in terms of universe trapped in a false vacuum has also 
been considered in Ref. \cite{Axenides}.
Alternate models which explain dark energy in terms
of an axion model are proposed in Ref. \cite{Kim,Choi,Pedro}.  

\bigskip
{\bf Acknowledgements:} We thank B. Ananthanarayan and John Ralston
for useful discussions.

\end{document}